\documentclass[preprint,number]{elsarticle}
\usepackage{amssymb}
\usepackage{graphics}
\usepackage{lineno}
\journal{Nuclear Instruments and Methods in Physics Research A}
\newcommand{\affinfn}[2]{INFN Sezione di #1, #2, Italy.}
\newcommand{\affuni}[2]{Dipartimento di Fisica dell'Universit\`a #1, #2,
  Italy.} %
\begin{document}
\begin{frontmatter}
\title{Measurement of the neutron detection efficiency of a 80\% absorber - 20\%
  scintillating fibers calorimeter.}
\author[Frascati]{M.~Anelli}
\author[Frascati]{S.~Bertolucci}
\author[Roma1,infnRoma1]{C.~Bini\corref{cor}} 
\ead{cesare.bini@roma1.infn.it}
\cortext[cor]{Corresponding author, Dipartimento di Fisica, Sapienza
  Universit\`a di Roma, P.le A.Moro, 2 I-00185 Roma, Italy;
  Tel. +390649914266, Fax +39064957697} 
\author[infnRoma3]{P.~Branchini} 
\author[Frascati]{G.~Corradi} 
\author[Frascati]{C.~Curceanu} 
\author[Roma1,infnRoma1]{G.~De Zorzi} 
\author[Roma1,infnRoma1]{A.~Di Domenico} 
\author[Roma3,infnRoma3]{B.~Di Micco} 
\author[cnao]{A.~Ferrari}
\author[Roma1,infnRoma1]{S.~Fiore}
\author[Roma1,infnRoma1]{P.~Gauzzi\corref{cor}}
\ead{paolo.gauzzi@roma1.infn.it}
\author[Frascati]{S.~Giovannella} 
\author[Frascati]{F.~Happacher} 
\author[Frascati,bucharest]{M.~Iliescu} 
\author[Frascati]{A.~Luc\`a}
\author[Frascati]{M.~Martini} 
\author[Frascati]{S.~Miscetti}
\author[Roma3,infnRoma3]{F.~Nguyen} 
\author[infnRoma3]{A.~Passeri}
\author[tsl]{A.~V.~Prokofiev} 
\author[Frascati]{I.~Sarra} 
\author[Frascati]{B.~Sciascia}
\author[Frascati,bucharest]{F.~Sirghi}
\author[Frascati]{D.~Tagnani}
\address[Frascati]{Laboratori Nazionali di Frascati dell'INFN, Via E.Fermi
  40, I-00044 Frascati, Italy.} 
\address[Roma1]{Dipartimento di Fisica, Sapienza Universit\`a di
  Roma, P.le A.Moro, 2 I-00185 Roma, Italy.}
\address[infnRoma1]{\affinfn{Roma}{P.le A.Moro, 2 I-00185 Roma}}
\address[infnRoma3]{\affinfn{Roma Tre}{Via della Vasca Navale, 84 I-00146
    Roma}}
\address[Roma3]{\affuni{``Roma Tre''}{Via della Vasca Navale, 84
    I-00146 Roma}}
\address[cnao]{Institute of Safety Research and Institute of Radiation
  Physics, 
  Forschungszentrum Dresden-Rossendorf, PF 510119, 01314 Dresden, Germany.}
\address[bucharest]{``Horia Hulubei'' National Institute of Physics and
  Nuclear Engineering, Str. Atomistilor no. 407, P.O. Box MG-6
  Bucharest-Magurele, Romania.} 
\address[tsl]{The Svedberg Laboratory, Uppsala University, Box 533, S-751
  21 Uppsala, Sweden.}

\begin{abstract}
The neutron detection efficiency of a sampling calorimeter made of 
1 mm diameter scintillating fibers embedded in a 
lead/bismuth structure has been measured at the neutron beam of The Svedberg
Laboratory at Uppsala. A significant enhancement of the detection efficiency 
with respect to a bulk organic scintillator detector with 
the same thickness is observed. 
\end{abstract}

\begin{keyword}
Neutron detection \sep Calorimetry \sep Scintillating fibers

\end{keyword}

\end{frontmatter}


\section{Introduction}
Neutrons of kinetic energies ranging between few and few
hundreds MeV are normally detected using organic
scintillators, the high concentration of hydrogen atoms providing
the proton target where the neutrons are elastically scattered. In this
energy range the efficiency 
of scintillator slabs smoothly depends on
the neutron kinetic energies and is typically of 1\% for every cm
of scintillator thickness \cite{scint}.\\
It has been suggested that the insertion of a high Z
material in a structure of organic scintillators could significantly  
enhance the neutron detection efficiency,
due to the abundant production of secondary particles in inelastic
interactions of the neutrons with the high Z material \cite{Baumann}. 
Compact detectors for neutrons of these energies are required by the
experiments proposed at the $e^+e^-$ collider DA$\Phi$NE of the
Frascati Laboratories, aiming to measure the neutron time-like form 
factors \cite{KLOE2}
and the low energy charged kaon interactions on nuclei \cite{AMADEUS}. 
In view of these projects, recently the neutron detection efficiency of a
lead-scintillating fibers calorimeter where the percentage in
volume of the active medium is about 50\% (the KLOE calorimeter \cite{kloe})
has been measured. An efficiency enhancement of a factor about 3 has
been measured with respect to a NE-110 bulk scintillator with an equivalent
scintillator thickness exposed to the same beam \cite{klone}.\\

In this paper we present the results of the experimental 
study of the response and of the detection efficiency to neutrons of
a different heterogeneous calorimeter module which is still based 
on scintillating fibers but with a lower 
percentage of active medium, about 20\% in volume. Note that lead
scintillating fibers calorimeters with this percentage of scintillator,
allow also to obtain the best compensation for hadron calorimetry \cite{Wigmans}.

\section{The calorimeter}
The calorimeter module \cite{fib}, 
32 cm in length and (7.5 $\times$ 7.5) cm$^2$ in
cross-section was built with a fusion technique: a high density alloy
(9.9 g/cm$^3$; alloy composition: 52.5\% Bi + 32.0\% Pb + 15.5\% Sn, in
weight) with low temperature melting point encloses an array of thin
stainless steel tubes 100 $\mu m$ thick, each containing a 1 mm diameter
scintillating fiber.  
The fibers, Kuraray SCSF-81, run parallel along
the module length and are positioned in the transverse plane at the
vertices of squares of 2.1 mm side. 
The overall structure is characterized by a scintillator percentage in
volume of 19.5\%.  
At each module edge the fibers are grouped together in two bundles
directly connected to Hamamatsu fine-mesh R5946 1.5' photomultiplier 
tubes (PM). The gains
of the four PMs have been adjusted in such a way to have the
same response to a minimum ionizing particle (mip) for each channel.  
Each PM signal is split and discriminated: charge and time distributions
are obtained through the KLOE ADC and TDC modules \cite{kloe}. 
The analog sums of the two signals from 
each side of the module provide the auto-trigger: 
the trigger requires both analog
sums to be larger than a given threshold V$_{th}$ that can be varied in a 
wide range.

In order to compare the efficiency extracted from these
measurements with those from different detectors, we express the threshold
in equivalent 
electron energy $E_{th}$ (MeV el.eq.). The correspondence between $V_{th}$ and
$E_{th}$ is given by:
\begin{equation}
E_{th}({\rm MeV~el.eq.})=V_{th}({\rm mV})\times{{\lambda({\rm count/mV})\delta E
    ({\rm MeV})}\over{MIP({\rm count}) (e/mip)}} 
\end{equation}
Here $\lambda$ is the conversion factor from ADC
counts to mV; $MIP$ is the peak value in ADC counts mips crossing the
calorimeter with the same direction of the neutrons. This mip sample has been 
obtained in a cosmic ray run with the calorimeter rotated by 90$^o$ around
the fiber direction triggered by two small scintillators. 
$\delta E=80$~MeV is the calculated energy release by a mip in the
calorimeter module; $(e/mip)$ is the ratio between the response to a 80 MeV
electron and to a mip \cite{Wigmans}.
Our $(e/mip)$ estimate is based on the cosmic ray run (see above) compared to 
runs with electrons at the Beam Test Facility (BTF) of the Frascati 
Laboratories. 
We obtain a value of $(e/mip)$ compatible with 1 within 15\%. 
Due to the small longitudinal size of our calorimeter, we correct for the
leakage affecting the electron energy measurement. 
On the other hand, a value of $(e/mip)=0.72\pm0.03$ is obtained in
Ref.\cite{Acosta} for a scintillating fiber calorimeter with a filling
factor of about 20\% but with a different absorber composition 
and for much higher energies (above 4 GeV). 
In the following, we consider the interval between the two values 
of $(e/mip)$ as a systematic uncertainties of the absolute electron
equivalent energy scale.\\ 
The threshold can also be expressed in number of photo-electrons (p.e.)
using the measured values of the PM gains. 
We obtain a correspondence approximately of 1 p.e. / MeV el.eq. We have
explored a threshold range between 6 and 120 MeV per side.\\

\section{Experimental set-up}
\begin{figure}[htb]
\begin{center}
\includegraphics[width=.8\textwidth]{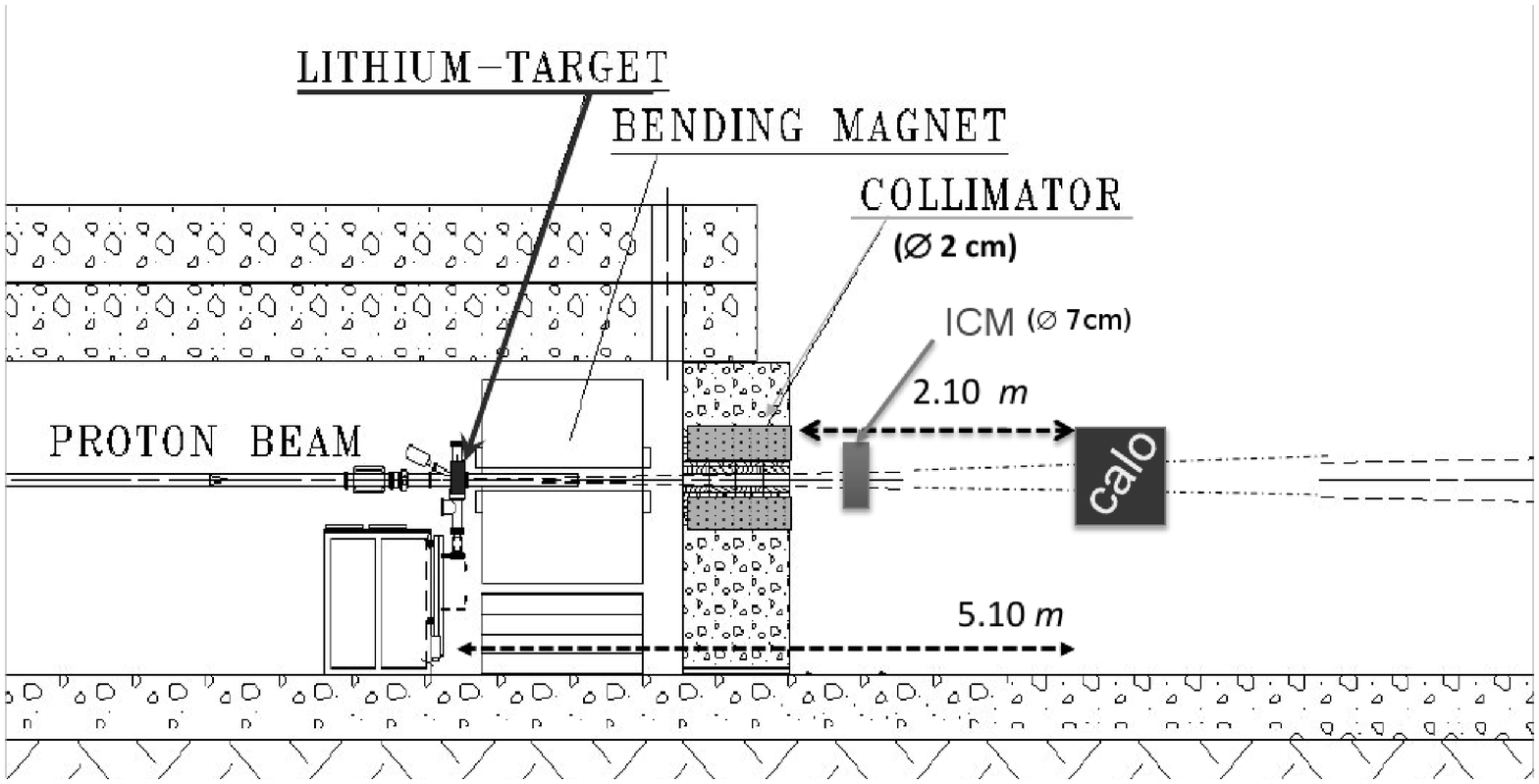} \\
\includegraphics[width=.6\textwidth]{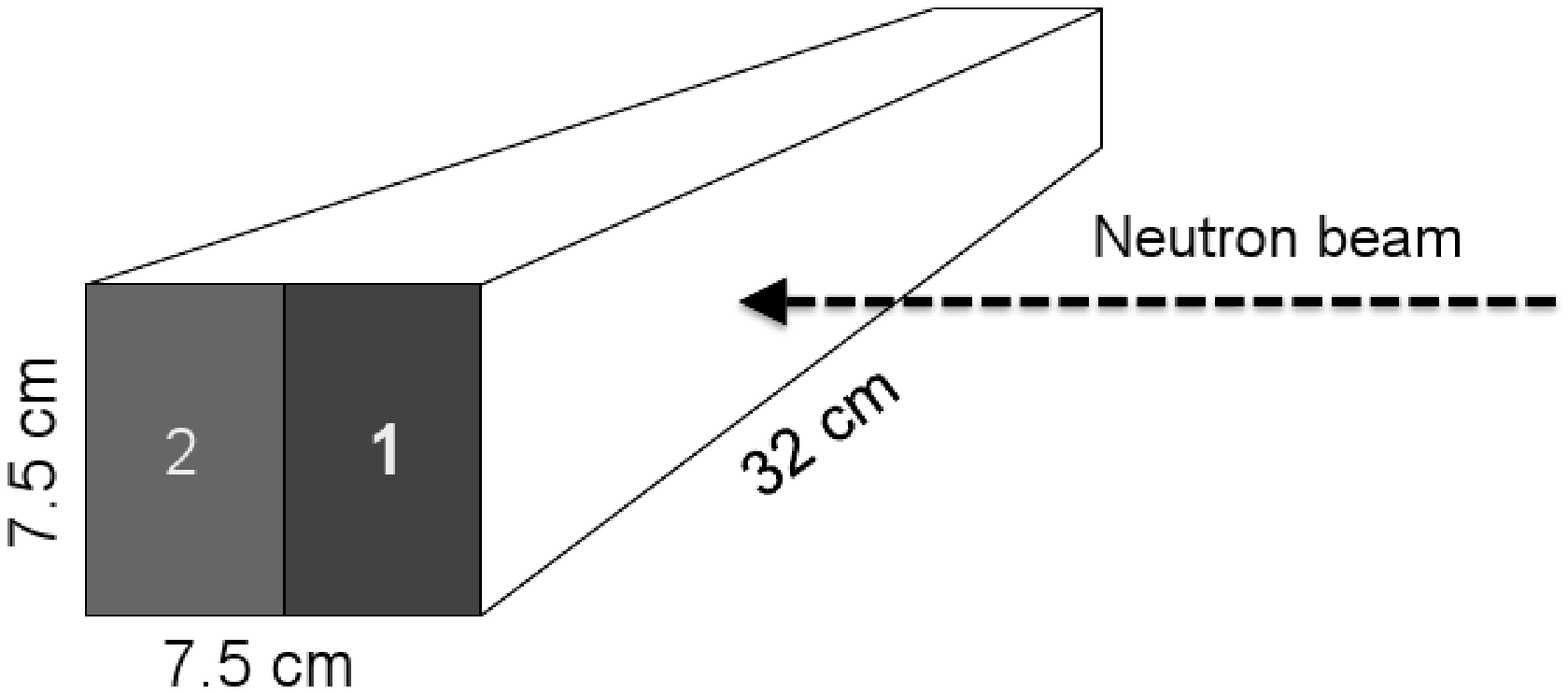}
\end{center}
\caption{TSL experimental lay-out: (top) calorimeter position on the TSL
  beam-line; (bottom) enlarged view of the calorimeter.}
\label{layout}
\end{figure}

We have exposed the calorimeter module to the neutron beam of The Svedberg
Laboratory (TSL) in Uppsala University \cite{TSL}. 
A proton beam of 178.7 MeV kinetic energy is sent to a 23.5 mm thick $^7$Li
target at a frequency of about 20 MHz. A bending magnet and an iron collimator block about 1 m long 
with a 2 cm diameter cylindrical hole allow to select a neutron beam with
negligible contamination. Fig.\ref{layout} shows the experimental layout. 

\begin{figure}[htb]
\begin{center}
\includegraphics[width=.6\textwidth]{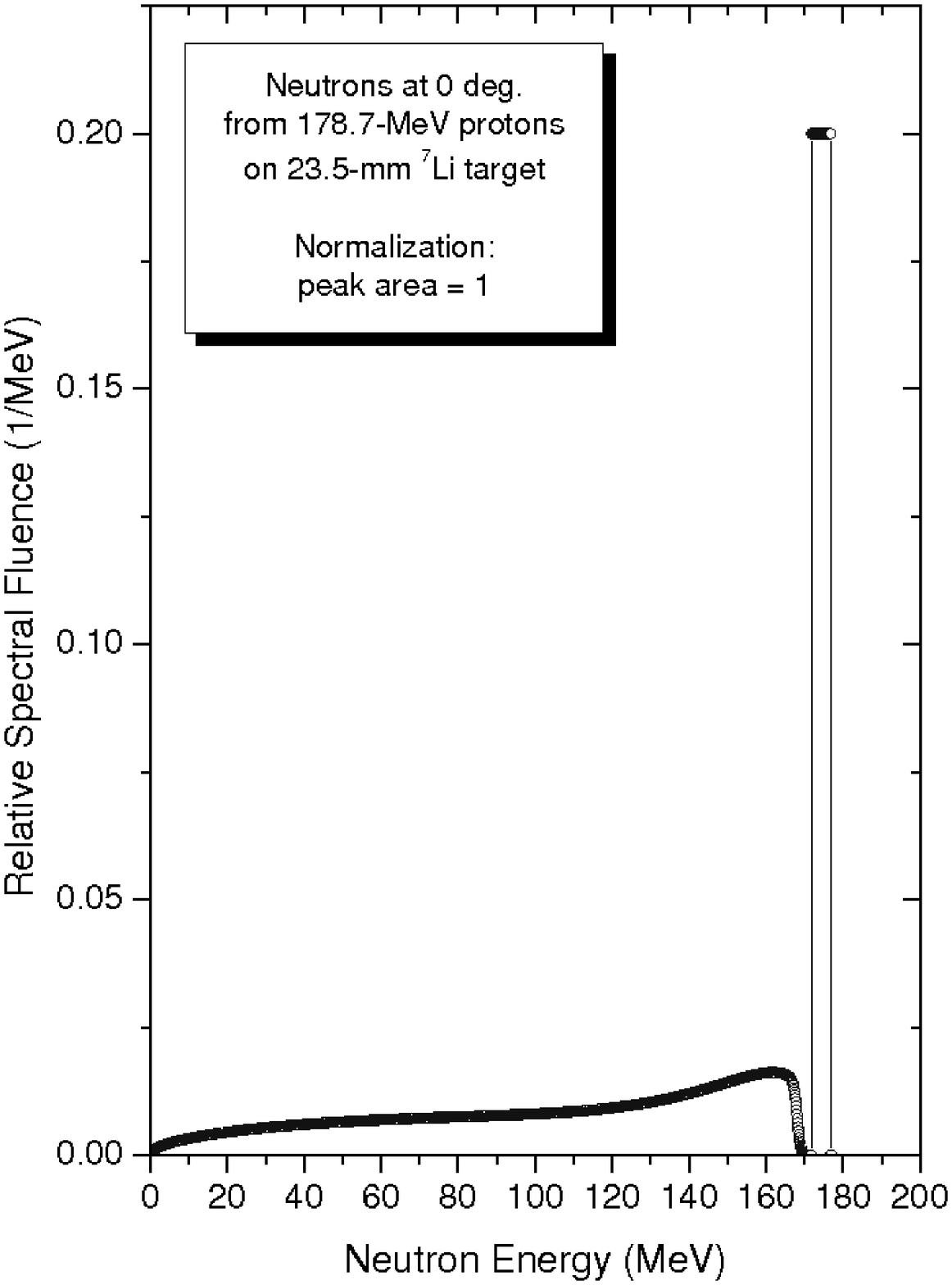}
\end{center}
\caption{TSL expected neutron kinetic energy spectrum out 
of the collimator.}
\label{spectrum}
\end{figure}

The kinetic energy 
spectrum of the neutrons emerging from the hole is shown in
Fig.\ref{spectrum}.
It is characterized by the quasi-elastic peak at 174 MeV
and by a tail extending to very low energies, the integral of the
quasi-elastic peak being about 40\% of the total spectrum. 
The neutron flux at the collimator exit is measured online with an Ionization
Chamber Monitor (ICM).
The ICM measurements are periodically 
calibrated with another monitor based on Thin-Film
Breakdown Counter in order to obtain the absolute neutron rate, that is
provided with an accuracy of about 15\%. 
The beam has two components: a bulk component with a Gaussian shape and
a larger halo which is 
not fully accounted for by the ICM. 
Therefore the total neutron rate has to be corrected in order
to properly normalize the absolute efficiency measurement.\\
A 20 MHz signal synchronous with the proton beam is provided to the
experiment and is used to strobe the calorimeter trigger signal 
thus acting as
start signal for the neutron time of flight measurement. 

\section{Results and discussion}
\begin{figure}[htb]
\begin{center}
\includegraphics[width=.49\textwidth]{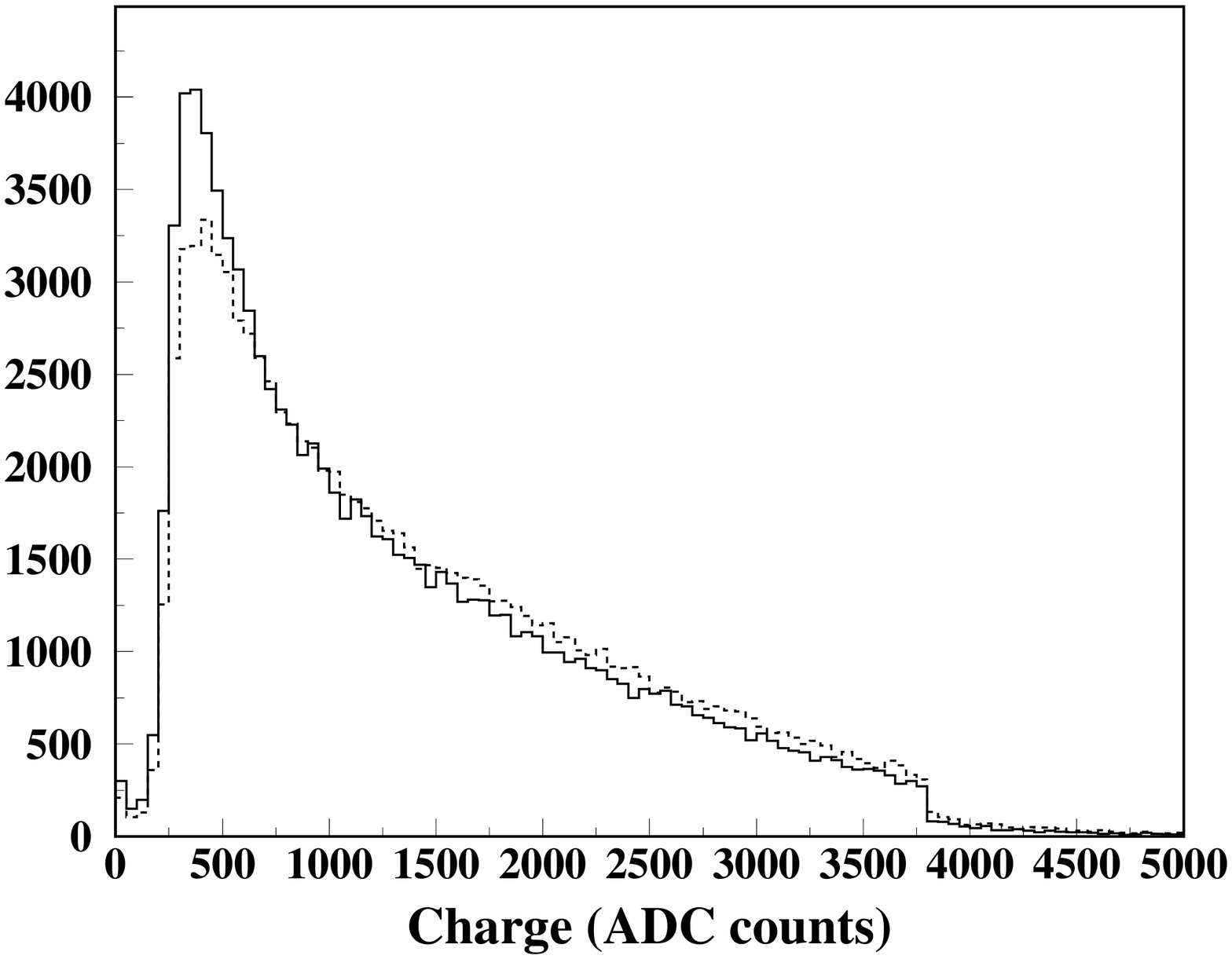}
\includegraphics[width=.49\textwidth]{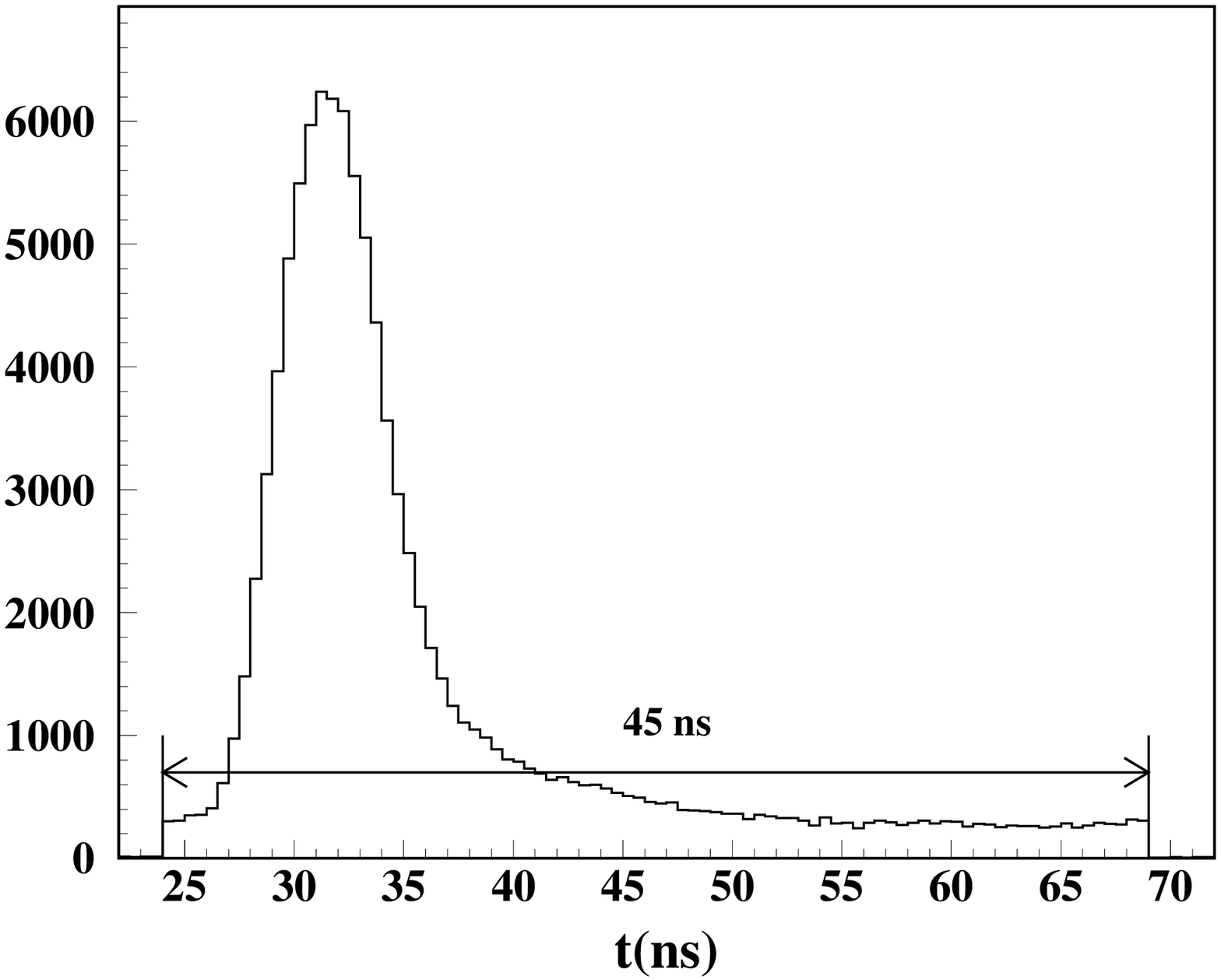}
\end{center}
\caption{(left) Charge spectrum of side A (solid histogram) and side B 
(dashed histogram) for the run with a threshold of 36 MeV. (right) Charge
 weighted time spectrum for the same run. The fixed time width of 45 ns due to 
the 20 MHz structure of the start signal can be clearly seen.}
\label{enetime}
\end{figure}

Charge and time spectra of the calorimeter are shown
in Fig.\ref{enetime} for a typical run with $E_{th}$=36 MeV
el.eq.. Continuous charge spectra clearly cut at the threshold
value
and characterized by a tail extending up to the ADC saturation value
corresponding to about 450 MeV el.eq. are obtained. 
The time spectrum is correlated to the neutron kinetic energy
spectrum, the large peak corresponding to the 174 MeV neutron peak
and the large time tail corresponding to lower kinetic energies.
The time spectrum is used 
to estimate the efficiency dependence on the neutron energy
as shown in the following. 
However, such correlation is reduced by the time resolution
and, to a minor extent, by a small amount of re-phasing effect due to the 20 MHz structure of the start
signal. 
This latter effect results in a fixed time spectrum width of 45 ns, and 
implies that the time of flight of very low energy neutrons can overlap
with the one of 174 MeV neutrons.  

The overall detection efficiency of the calorimeter is defined as:
\begin{equation}
\epsilon={{r_{calo}}\over{r_{ICM}}}\times{{f_{ICM}}\over{A}}
\label{eq:effi}
\end{equation}
where: 
\begin{itemize}
\item{$r_{calo}$ is the calorimeter trigger rate, the accidental trigger
  rate being negligible;} 
\item{$r_{ICM}$ is the neutron rate measured by ICM; }
\item{$f_{ICM}$ is the ratio between the integral of the bulk neutron
  distribution and the overall neutron rate impinging  
  on the calorimeter;}
\item{$A$ is the acceptance of the calorimeter for bulk neutrons.} 
\end{itemize}

\begin{figure}[hbt]
\begin{tabular}{cc}
  \includegraphics[width=.5\textwidth]{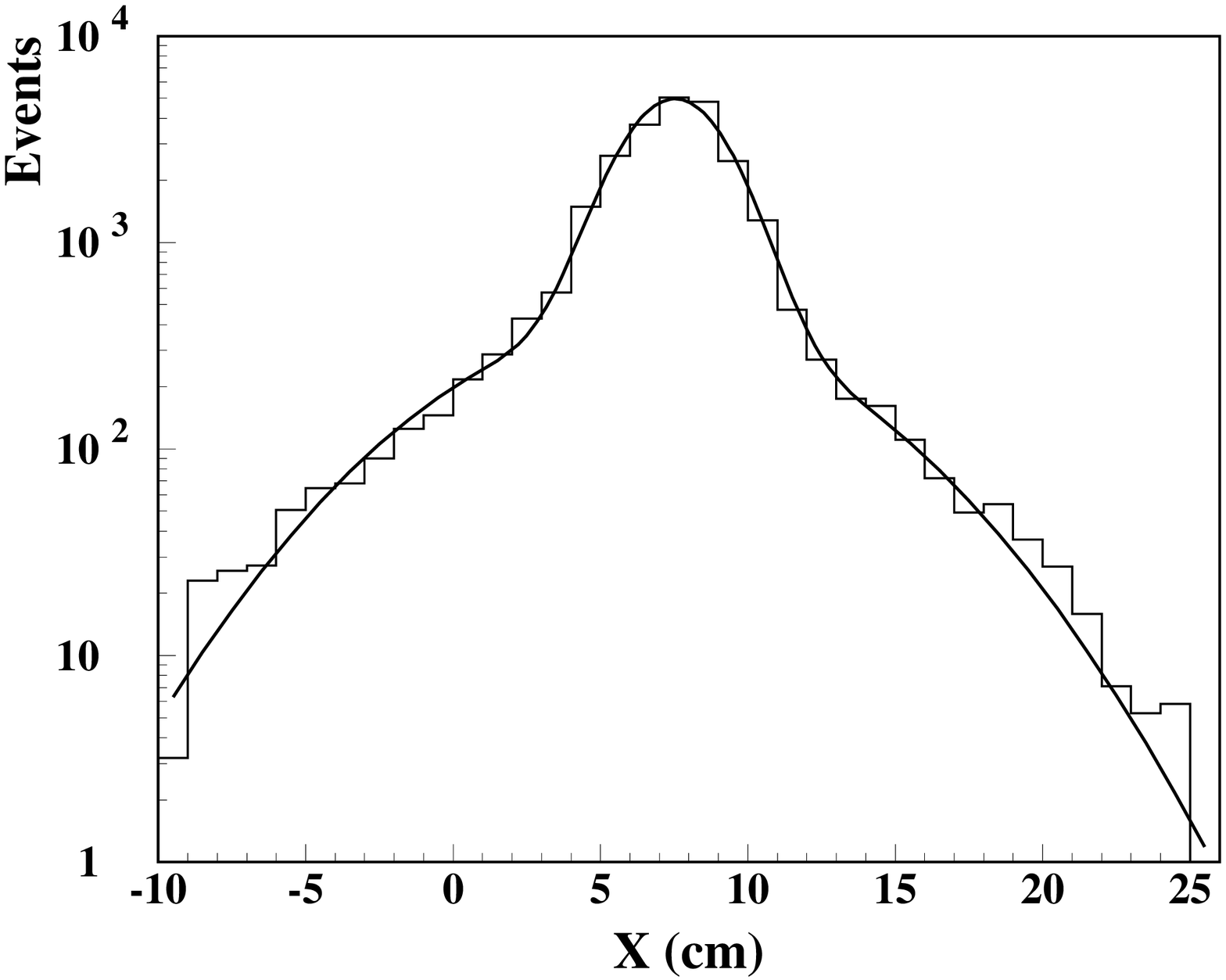}
  \includegraphics[width=.5\textwidth]{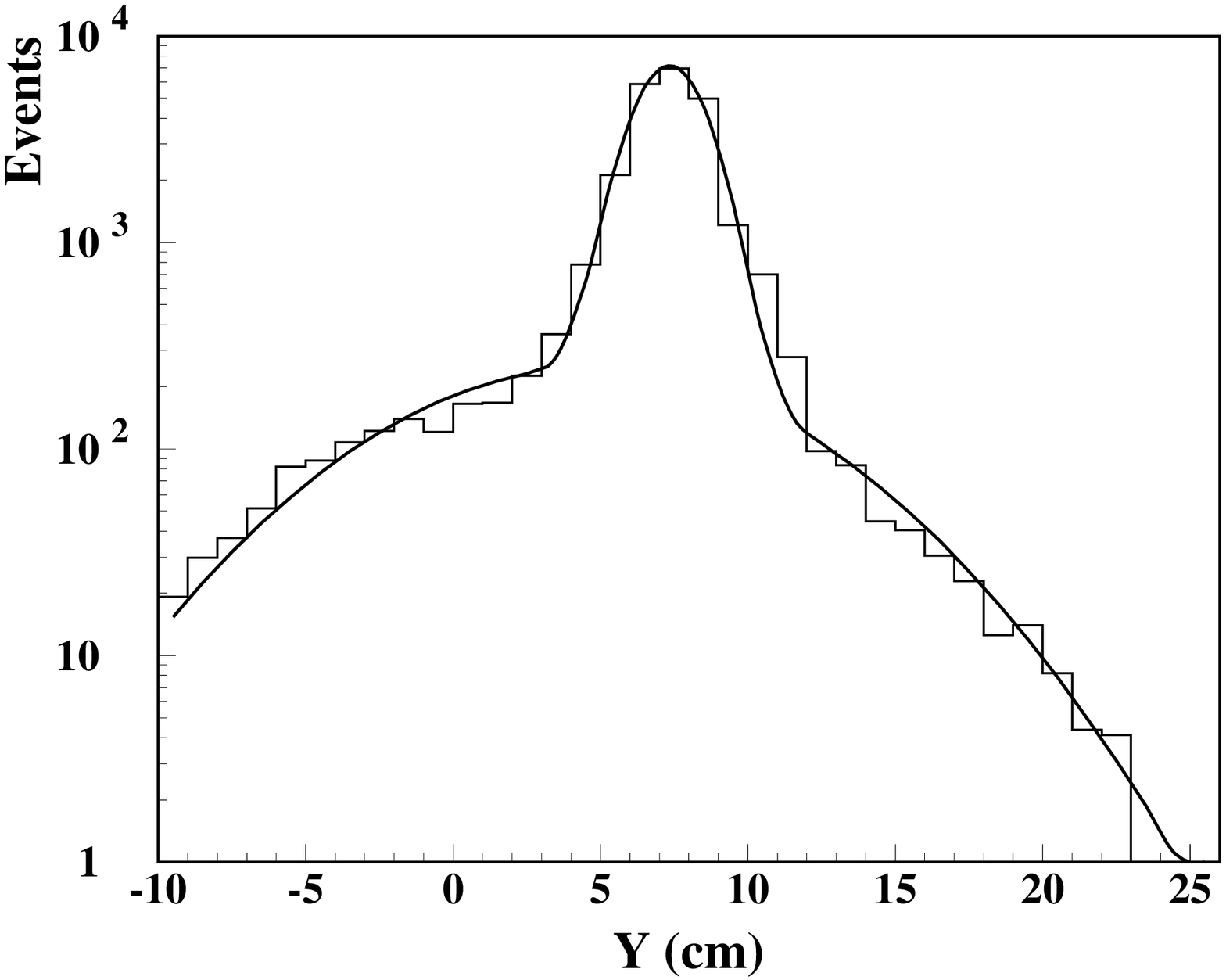}
\end{tabular} 
\caption{Distribution of the counts in the BPM scintillators in
  the horizontal (let plot) and vertical (right plot) directions. The
  curves superimposed are bi-gaussian fits.  }
\label{fig:bpm}
\end{figure}

The value of $f_{ICM}$ depends on the beam shape.
We have measured the beam shape 
in both horizontal (X) and vertical (Y) directions, by using a Beam
Profile Monitor (BPM).
The BPM consists of two orthogonal hodoscopes, each made of 16
scintillator slabs, 1 cm wide and 0.5 cm thick, readout by 
Hamamatsu H8711 4$\times$ 4 multianode PM's through Saint Gobain BCF92
wavelength shifter fibers.
The BPM threshold for neutron detection has been estimated to be smaller than
10 MeV equivalent electron energy.
The beam profile has been measured at different distances from the
collimator exit.
The presence of a bulk component pointing to the $^7$Li target, surrounded by a
larger and non pointing halo, has been observed.
The X and Y distributions at the position of the calorimeter module are
shown in Fig.\ref{fig:bpm}: the RMS width of the bulk component is 1.6 cm in X
and 1.2 cm in Y, while the halo has a width of about 6 cm in both
directions and its center is slightly displaced with respect to the bulk.  

The beam shape at the position of the calorimeter has been then parametrized 
with a bidimensional function reproducing the X and Y distributions of
Fig.\ref{fig:bpm} and a value $f_{ICM}= 0.92\pm 0.03$ has been obtained by
integrating that beam shape over the sensitive surface of the calorimeter. 
The uncertainty takes into account the stability of the measurement with
respect to the event selection, and a difference in detection
efficiency for bulk and halo neutrons due to their possible different energy
composition.  
Since the bulk neutron distribution is fully contained in the calorimeter
sensitive surface the acceptance $A$ is  equal to 1.

\begin{figure}[htb]
\begin{center}
\includegraphics[width=.6\textwidth]{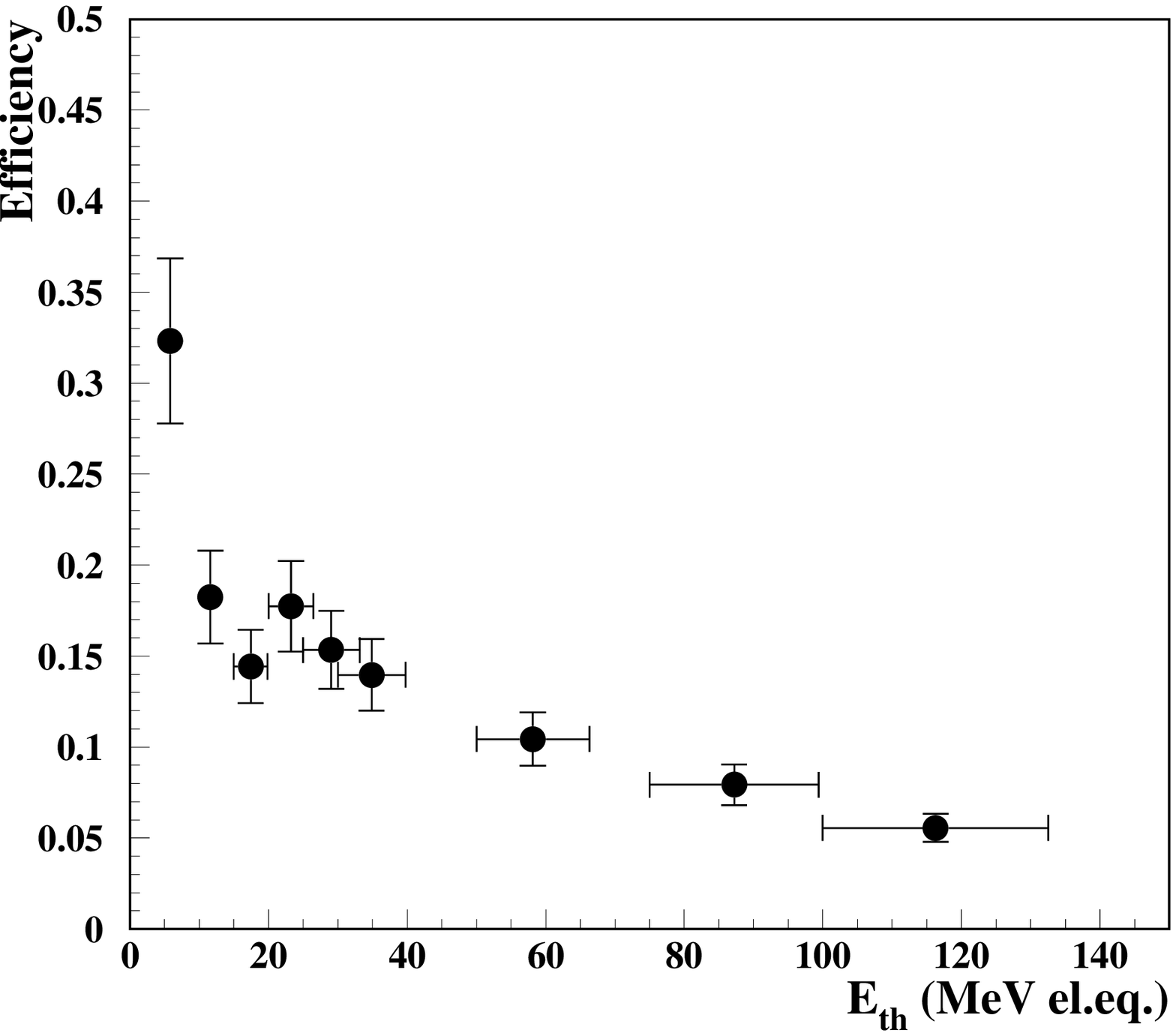}
\end{center}
\caption{Neutron overall efficiency as a function of the threshold
  expressed in MeV of electron equivalent for the calorimeter. }
\label{fig:results}
\end{figure}

The overall detection efficiency has been measured for different values of
the trigger threshold V$_{th}$. 
Table \ref{tab:results} and Fig.\ref{fig:results} show the results of the 
measurement.
\begin{table}[htb]
    \caption{Results of the measurement: calorimeter rates and beam rates
    as a function of the central value of the threshold interval. 
The overall efficiency is evaluated according
    to eq.(\ref{eq:effi}).}
\begin{center}
    \begin{tabular}{|c|c|c|c|} \hline
 E$_{th}$(MeV el.eq.) & $r_{calo}$(kHz) & $r_{ICM}$(kHz) & $\epsilon$ \\
\hline
6    & 6.64 & 18.9  & 0.323 \\
12   & 4.94 & 24.9  & 0.184 \\
18   & 4.22 & 26.9  & 0.144 \\
24   & 3.53 & 18.3  & 0.178 \\
30   & 3.15 & 18.9  & 0.154 \\
36   & 2.84 & 18.7  & 0.140 \\
60   & 2.27 & 20.0  & 0.106 \\
89   & 1.82 & 21.1  & 0.079 \\
119   & 1.53 & 25.3  & 0.056 \\
\hline
   \end{tabular}
  \end{center}
  \label{tab:results}
\end{table}

The overall uncertainty of about 14\% is dominated by the 
uncertainty on $r_{ICM}$. The systematic error
on the threshold absolute scale due to the uncertainty on the $(e/mip)$ 
value (see above) has also been considered.\\
Table \ref{tab:comparison} compares the measured overall efficiency with the one
obtained in the same beam by a NE-110 bulk scintillator and by the KLOE
calorimeter \cite{klone}.\\ 
\begin{table}[htb]
    \caption{Overall efficiency per unit of scintillator thickness
    ($\eta_A$) and per unit of overall detector thickness ($\eta_B$) for
    NE-110 bulk scintillator \cite{klone}, KLOE calorimeter (50\% fibers
    and 50\% absorber)\cite{klone} and for the calorimeter module (20\%
    fibers and 80\% absorber, this work). All measurements have been done
    for neutrons having the spectrum of Fig.\ref{spectrum} and with the
    same threshold of about 10 MeV.el.eq..} 
\begin{center}
    \begin{tabular}{|c|c|c|} \hline
  & $\eta_A$ (\%/cm) & $\eta_B$ (\%/cm)  \\
\hline
NE-110 & 0.6 & 0.6 \\
KLOE & 1.9 & 0.9 \\
this calorimeter & 12 & 2.4 \\
\hline
   \end{tabular}
  \end{center}
  \label{tab:comparison}
\end{table}
The three detectors are operated with the same
10 MeV el.eq. threshold and are exposed to the same beam with the spectrum
shown in Fig.\ref{spectrum}\footnote{In the comparison between the KLOE calorimeter and the
calorimeter considered in this paper, the different
absorber composition has to be considered: 95\% Pb - 5\% Bi (KLOE) and 32\% Pb - 52.5\% Bi -
15.5\% Sn (this calorimeter).}.
Two normalized efficiencies are defined: $\eta_{A}$ is the
efficiency normalized to the average thickness of active material (to
compare with bulk scintillator), $\eta_{B}$ is the
efficiency normalized to the overall detector thickness.\\ 
In order to estimate the kinetic energy dependence of the
efficiency we have fit the measured time spectrum shown in
Fig.\ref{enetime}(b) corresponding to the sample taken with a 
36 MeV el.eq. threshold. The fitting function is given by the
neutron time
of flight spectrum obtained from the kinetic energy one shown in
Fig.\ref{spectrum} through the relation
\begin{equation}
t={{z}\over{c\sqrt{1-\big({{M_n}\over{K_n+M_n}})^2}}}
\end{equation}
where $z$=510 cm is the distance of the calorimeter from the proton target,
$M_n$ 
and $K_n$ are the mass and kinetic energy of the neutron respectively.  
\begin{figure}[htb]
\begin{center}
\includegraphics[width=.49\textwidth]{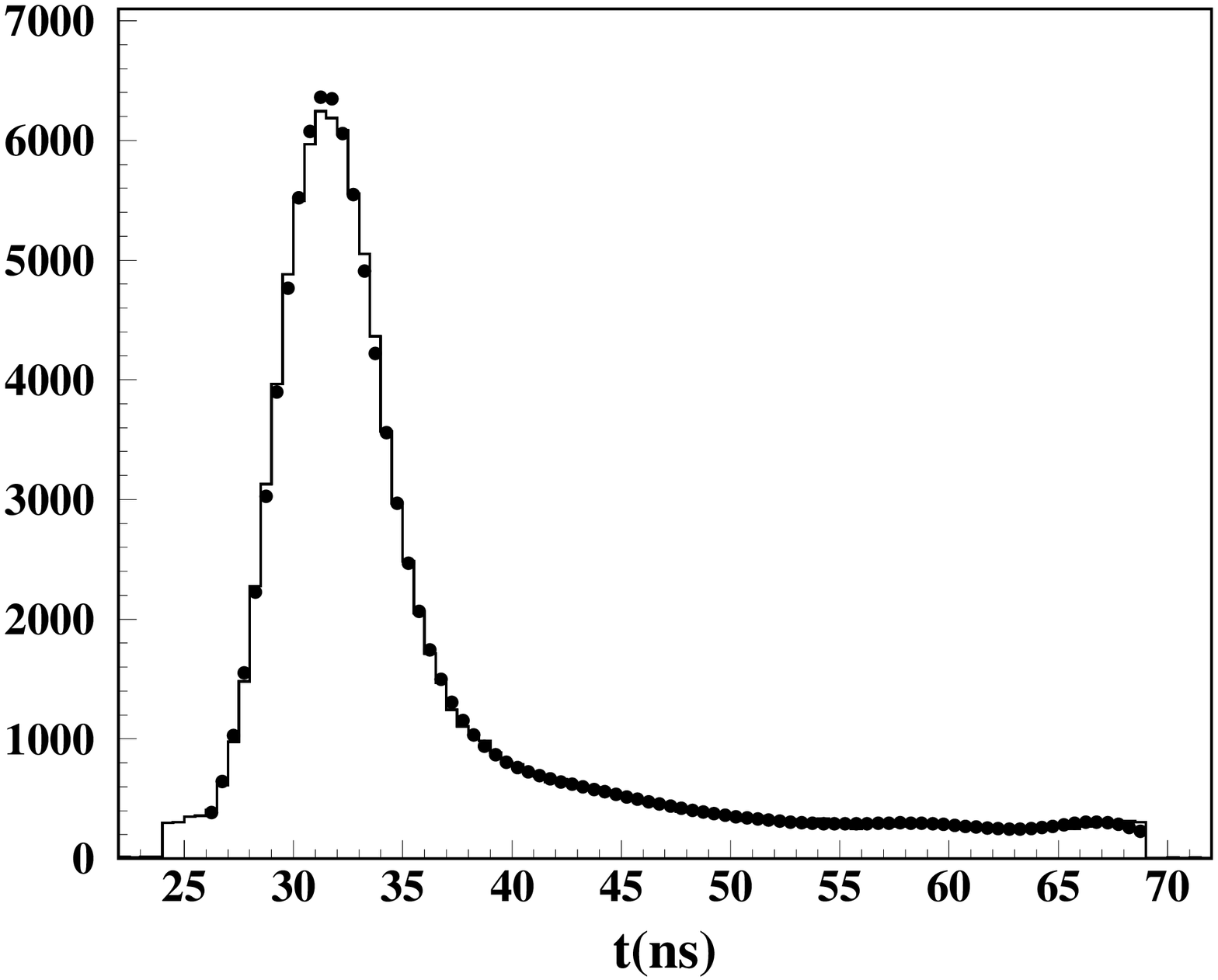}
\includegraphics[width=.49\textwidth]{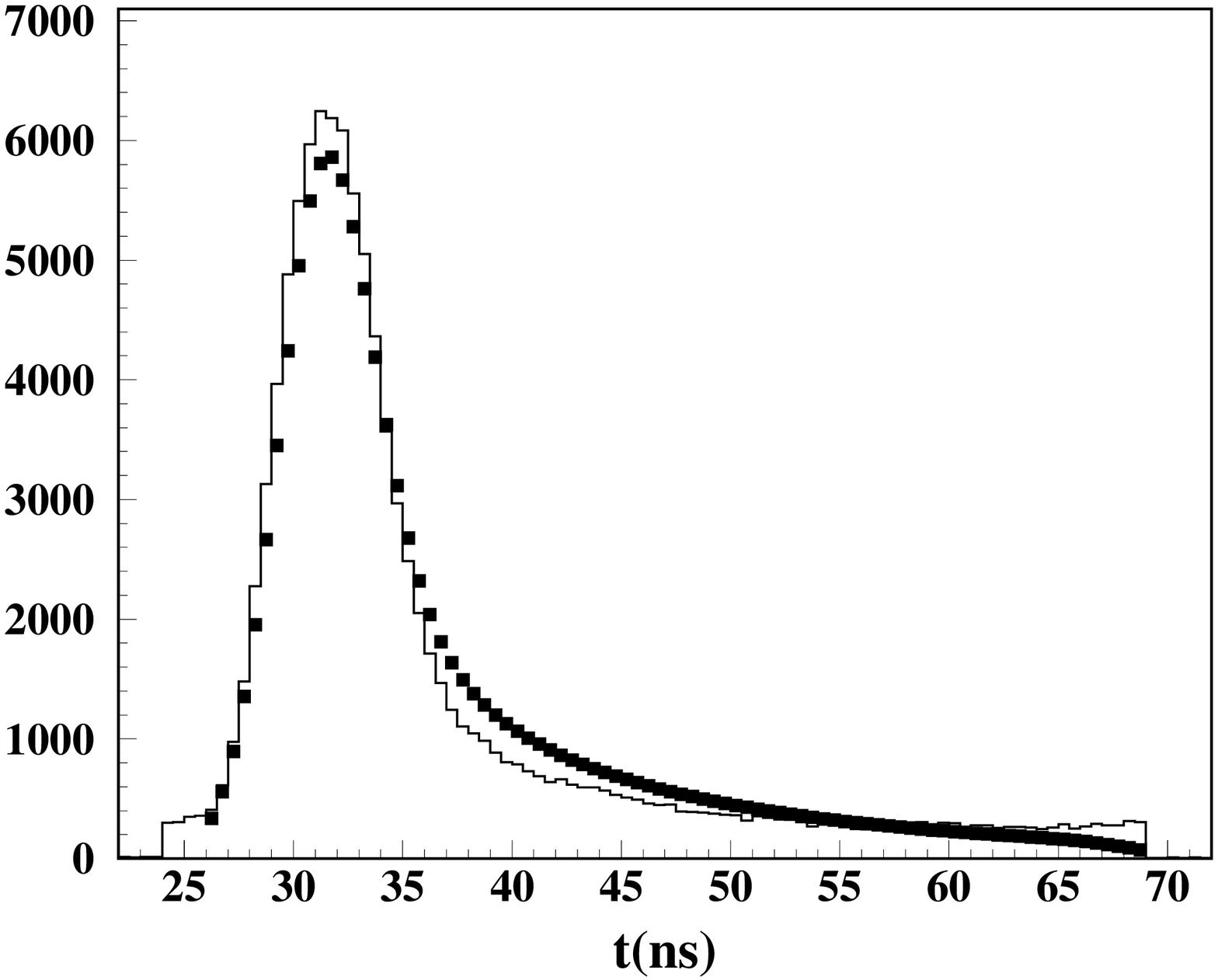}
\end{center}
\caption{(left) Time spectrum (solid histogram) with superimposed the fit 
result (points) and (right) the same spectrum with superimposed 
the expectation in case of uniform efficiency (squares). }
\label{fit}
\end{figure}

The spectrum has been 
convoluted with a time resolution function and multiplied by an
efficiency profile $\epsilon(t)$. The time of flight spectrum takes into account
the small percentage (of the order of 3\%) of rephased low energy
neutrons.
The time resolution is assumed to be gaussian
with a time independent width, and the efficiency is either binned in eight
time bins or is parametrized as a 4$^{th}$
order Cebychev polynomial. Free parameters are: the width 
of the time resolution and the eight efficiencies or the five 
coefficients of the Cebychev polynomials.
$\epsilon(t)$ is constrained to give an average
efficiency equal to the overall efficiency measured as shown above.\\
Fig.\ref{fit} shows the experimental spectrum superimposed to the fit
result (left plot) and also to the expected
one in case of uniform efficiency (right plot). A uniform efficiency
in the range 40-180 MeV does not adequately fit the data. 
\begin{figure}[hbt]
\begin{center}
\includegraphics[width=.6\textwidth]{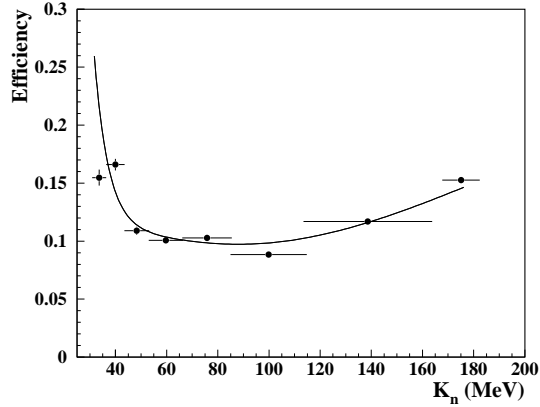}
\end{center}
\caption{Efficiency resulting from the fit as a function of 
the neutron kinetic energy $K_n$: the points are the results for the 8 bins fit
and the solid curve the one for the 4$^{th}$-order Cebychev polynomials.   }
\label{fit_bis}
\end{figure}

The efficiency resulting
from the fit is expressed as a function of the kinetic energy
in Fig.\ref{fit_bis} both for the binned and for the polynomial fit. 
A time resolution value of 2 ns is obtained in both cases. 
The data indicate an increase of efficiency both at
high and low kinetic energies.  

\section{Conclusions.}
The results of the study presented here show that the fraction of passive
absorber used in a heterogeneous calorimeter plays an important role in the
neutron detection. 
This also translates in a large enhancement factor with respect to a bulk
scintillator detector. \\
A first indication of an efficiency dependence on
the kinetic energy is also obtained. 


\paragraph{Acknowledgements}

We recall that the calorimeter was built in Rome some years ago
with the skilful help of M.Bertino.
We thank M.Arpaia, G.Bisogni, A.Cassar\`a, A.Di Virgilio, U.Martini,
A.Olivieri and all the Mechanical LNF Division for the help in the 
setup of the detector. 
We also acknowledge M.Rossi for the transportation of the material from
LNF to TSL.
Moreover we want to warmly thank all the TSL staff for the help
during the data taking.

\end{document}